\documentclass[11pt,a4paper,twoside,fleqn]{article}
\usepackage[top=0.77in,left=1.58in,right=1.77in,bottom=3.52in]{geometry}
\usepackage{amsmath,amsfonts,amssymb}
\usepackage{graphicx}
\usepackage{mathptmx}
\usepackage[margin=10pt,font=small,labelfont=bf,labelsep=quad]{caption}
\usepackage{avant}
\usepackage{fancyhdr}
\usepackage{enumitem}
\usepackage[hang,flushmargin]{footmisc} 
\usepackage[explicit]{titlesec}
\usepackage [running,mathlines] {lineno}
\usepackage{hyperref}
\hypersetup{colorlinks = true, linkcolor = blue, anchorcolor = red, citecolor = blue, filecolor = red, urlcolor = blue}

\newcommand\T{\rule{0pt}{2.6ex}}       
\newcommand\B{\rule[-1.2ex]{0pt}{0pt}} 

\usepackage{ITBjournal}

\journalID{
ITB J. ................... Vol. XX ..., No. X, 20XX, XX-XX}

\PublicationDate
{DD-MM-YYYY
}
{DD-MM-YYYY 
}
{DD-MM-YYYY 
}

\newcommand{\titlehead}{Properties of the Environment of Galaxies in Clusters of Galaxies}
\newcommand{\authorhead}{Premadi, P. W., Nugroho, D. H., \& Jaelani, A. T. 
}
\makeheader

\begin{document}


\begin{center}
\thispagestyle{fancy}
\title{Properties of the Environment of Galaxies in Clusters of Galaxies CL 0024$+$1654 and RX J0152.7$-$1357}
\author{Premana W. Premadi${}^{1*}$, Dading H. Nugroho${}^{1\dagger}$, \& Anton T. Jaelani${}^{1,2}$}
${}^1$Astronomy Research Group and Bosscha Observatory, FMIPA ITB, \\
Jalan Ganesha 10, Bandung 40132, Indonesia\\
${}^2$Department of Physics, Kindai University, 3-4-1 Kowakae, Higashi-Osaka, Osaka 577-8502, Japan\\
*Email: premadi@as.itb.ac.id\\
${}^{\dagger}$Deceased
\end{center}

\begin{abstract}
We report the results of combined analyses of X-ray and optical data of two galaxy clusters, CL 0024$+$1654 and RX J0152.7$-$1357 at redshift $z = 0.395$ and $z = 0.830$, respectively, which offer a holistic physical description of the two clusters. Our X-ray analysis yields temperature and density profile of the gas in the intra-cluster medium (ICM). Using optical photometric and spectroscopic data, complemented with mass distribution from gravitational lensing study, we investigate any possible correlation between the physical properties of the galaxy members, i.e., their color, morphology, and star formation rate (SFR) with their environments. We quantify the properties of the environment of each galaxy by galaxy number density, ICM temperature, and mass density. Although our result shows that the two clusters exhibit a weaker correlation compared to relaxed clusters, it still confirms the significant effect of the ICM on the SFRs in the galaxies. The closer relation between galaxy physical property and the condition of its immediate environment found in this work indicates the locality of galaxy evolution, even within a larger bound system such as cluster. Various physical mechanisms are suggested to explain the relation between the properties of galaxies and their environment.
\end{abstract}


\setcounter{footnote}{1}

\section{Introduction}
A cluster of galaxies is in general defined as a self-gravitating system of galaxies that may be identified observationally as a system exhibiting one or more of the following features: a significantly higher concentration of galaxies relative to the average distribution of large scale structure of the Universe; massive concentration(s) of X-ray emitting gas \cite{Sarazin+86}, often associated with a Sunyaev-Zel’dovich effect on the Cosmic Microwave Background Radiation \cite{Sehgal+11,Zeldovich+69}; a non-random gravitational lensing shear map of background galaxies \cite{Hoekstra+13}. Using computational simulation, clusters are identified as the largest gravitationally bound halo of dark matter of the large scale structure \cite{Springel+05,Vogelsberger+14}. A possible cluster is recognized as a crowding of galaxies within a span of a few Mpc on the projected plane with redshift representing that of the majority of galaxies in that crowd. This would be followed by estimation of the physical extension of the gravitational bound and inquiries of possible shared history, all that decides its status. In such a cluster, the proximity of one galaxy to another, and also to the ICM suggests the plausibility of environmental effect on galaxy evolution. With this reasoning, clusters of galaxies have become rich laboratories to study the formation and evolution of galaxies, and thus the evolution of the large-scale structure of the Universe.

The evolution of galaxies is complex due to the various physical processes involved. At the earliest stage of the galaxy formation, the initial condition, such as the  local density contrast and angular momentum distribution, sets off the process of protogalaxy collapse. The global dynamics of the Universe and the environmental feed-back must be taken into account at all times of the galaxy evolution. At later stages of the evolution, the process would become locally more complex, the physics become nonlinear requiring more parameters to describe it, and thus highly sophisticated computational technique and performance to solve the mathematical equations. To tackle such an inverse problem in astrophysics it is customary to use the interplay between two approaches: observational data analysis and theoretical (and computational) modelling. So far it has been the only way to probe the processes that might have taken place in galaxy evolution. We need to emphasize that there is no satisfying generic recipe for galaxy evolution that specify the roles of the initial condition and environmental effect at all scales and all times. The result is evident in the great variety that we see in galaxy features.

In general, the components of a cluster are identified as galaxies, unbound stars recognized as the intra-cluster light (ICL), ICM (e.g., hot gas), and dark matter. A supermassive black hole may reside at the center of a massive and relaxed cluster as the core of the Brightest Cluster Galaxy (BCG). BCGs are often found as giant massive galaxies. Blackholes are also known to be the dynamos of active galactic nucleus (AGN), which are cluster’s member giving rise to additional mechanisms to be considered in the cluster evolution. Each component may interact with one another, and there are suggested mechanisms for each interaction. The interactions are in general grouped into three, considering the participants: (1) interaction between ICM and galaxies, which involves the gaseous component of galaxies in direct contact with ICM, (2) interaction between the cluster potential and the galaxies, and (3) interaction between galaxies.

All those interactions may affect the internal structure of the involved galaxies indicated by the significantly different values of some physical parameters of the galaxies (e.g., SFR) compared to those of isolated galaxies which experience minimal environmental effect throughout their evolution.

The work presented here concentrates on identifying the properties of each galaxy belonging to the cluster, identifying the immediate environment of each of those galaxies, and studying any correlation between the properties of individual galaxies and their environments. The primary properties of each galaxy are taken to be their color, morphology, and SFR. The environment is characterised by its surrounding galaxy number density, mass density, and ICM temperature. We also examine any possible dependence of galaxy properties on the distance to the cluster center, such that this work may be regarded as comparative study to that of Dressler’s morphology-density relation versus morphology-distance relation \cite{Dressler+80}.

For the aforementioned purpose we choose to examine two well observed galaxy clusters, namely CL 0024$+$1654 and RX J0152.7$-$1357 at redshift $z = 0.395$ and $z = 0.830$, respectively (hereafter, CL0024 and RXJ0152). Throughout the paper we use cosmology with $\Omega_{\rm m}=0.3$, $\Omega_{\Lambda}=0.7$, and $H_0=73$ km s$^{-1}$ Mpc$^{-1}$, such that $z = 0.395$ corresponds with lookback time 4.07 Gyr, and $z = 0.830$ with 6.69 Gyr. Previous studies classified these two clusters as nonrelaxed: the currently observed CL0024 is produced by a head on merger of two clusters \cite{Jee+07,Jee+10,Wagner+18}, whereas RXJ0152 is a young cluster still accreting the smaller sub-clusters \cite{Demarco+05}. We choose to further study these two clusters to seek relation between the detail properties of the member galaxies, with the galaxies' immediate environment,  and also with the overall features of the clusters. Choosing these well studied clusters assists us in gauging the overall properties of the two clusters. This paper is organized as follows. In Section \ref{sec:data}, we describe the details of clusters data used in our analysis. The methods used for our analysis are given in Section \ref{sec:analysis}, and the results are presented in Section \ref{sec:results}. Finally, the discussion and conclusions of our analysis are given in Section \ref{sec:summary}.

\section{Data and Reduction}\label{sec:data}

Two types of data were used for this work: (1) a set of primary X-ray data from \textit{XMM-Newton} observation, (2) a set of secondary photometric and spectroscopic data as results of previous studies. There is a rich data collection available on CL0024, ranging from X-ray band to infrared band. 

Sources of data are as follows:  X-ray band data are from the \textit{Chandra Observatory}\footnote[1]{\label{note1}\href{https://cxc.harvard.edu/cda/}{https://cxc.harvard.edu/cda/}} \cite{Ota+04} and the \textit{XMM-Newton Observatory}\footnote[2]{\label{note2}\href{https://www.cosmos.esa.int/web/xmm-newton/xsa}{https://www.cosmos.esa.int/web/xmm-newton/xsa}} \cite{Zhang+05}; \textit{GALEX}\footnote[3]{\label{note3}\href{http://www.galex.caltech.edu/researcher/data.html}{http://www.galex.caltech.edu/researcher/data.html}} for ultraviolet band \cite{Moran+06}; optical data from the \textit{Canada-France-Hawaii Telescope} (CFHT)\footnote[4]{\label{note4}\href{https://www.cadc-ccda.hia-iha.nrc-cnrc.gc.ca/en/cfht/}{https://www.cadc-ccda.hia-iha.nrc-cnrc.gc.ca/en/cfht/}}, the \textit{William Herschel Telescope} (WHT)\footnote[5]{\label{note5}\href{http://www.ing.iac.es/astronomy/telescopes/wht/}{http://www.ing.iac.es/astronomy/telescopes/wht/}} \cite{Czoske+01}, and the \textit{Subaru Telescope}\footnote[6]{\label{note6}\href{https://subarutelescope.org/en/}{https://subarutelescope.org/en/}} \cite{Kodama+04}; the \textit{Infrared Space Observatory} (ISO) satellite\footnote[7]{\label{note7}\href{https://www.cosmos.esa.int/web/iso}{https://www.cosmos.esa.int/web/iso}} \cite{Coia+05} and the \textit{Spitzer Space Telescope}\footnote[8]{\label{note8}\href{https://www.spitzer.caltech.edu/}{https://www.spitzer.caltech.edu/}} \cite{Geach+06} for infrared band. RXJ0152 is one of the few observed high-redshift clusters with multi-band observations: \textit{Chandra X-ray} observation reported in \cite{Maughan+03} and \textit{XMM-Newton} in \cite{Maughan+06}; mass distribution from the Hubble observation using weak lensing analysis given in Jee \textit{et al.} in \cite{Jee+05}; and spectroscopic studies are given in Demarco \textit{et al.} in \cite{Demarco+05}, Homeier \textit{et al.} in \cite{Homeier+05}, and Jørgensen \textit{et al.} in \cite{Jorgensen+05}. The general properties of the two clusters are given in Table \ref{tab:tab1}, where $N_{\rm mem}$ means the number of galaxy member having magnitude $m$ in the range $m_{3}$ to $m_{3}+2$, following Abell's cluster richness criterion, where $m_{3}$ is the magnitude of the third brightest galaxy \cite{Abell58}; and $r_{\rm phys}$ means physical radius. 

\begin{table}[!htbp]
\centering
\caption{Summary of the two galaxy clusters properties.}
\fontsize{9pt}{9pt}\selectfont
\begin{tabular}{lllccc}
\hline \T
\textbf{Cluster}&\textbf{$\alpha$ (J2000)}&\textbf{$\delta$ (J2000)}&\textbf{$z$} &\textbf{$N_{\rm mem}$} & \textbf{$r_{\rm phys}$}\B\\
\hline\T
CL0024  & 00:26:35.7 & $+$17:09:43.1 & 0.395 & 503 & 1.08 Mpc\\
RXJ0152 & 01:52:41.4 & $-$13:59:20.1 & 0.830 & 123 & 1.51 Mpc\\
 \hline
\end{tabular}
\label{tab:tab1}
\end{table}

\subsection{X-ray}

The X-ray data for this analysis were taken from the \textit{XMM-Newton} public archive and all were obtained using the European Photon Imaging Camera (EPIC) in standard Full Frame (FF) mode and Extended Full Frame (EFF) mode for MOS and pn CCDs, respectively. For all detectors, the thin filter was used. CL0024 was observed on January 6$^{\rm th}$, 2001 for a total exposure time of 52.1 ks, 52.1 ks and 48.3 ks for MOS1, MOS2 and pn. RXJ0152 was observed on December 24$^{\rm th}$, 2002 for a total exposure time 50 ks.

\begin{figure}[!htp]
\centering
\includegraphics[width=\columnwidth]{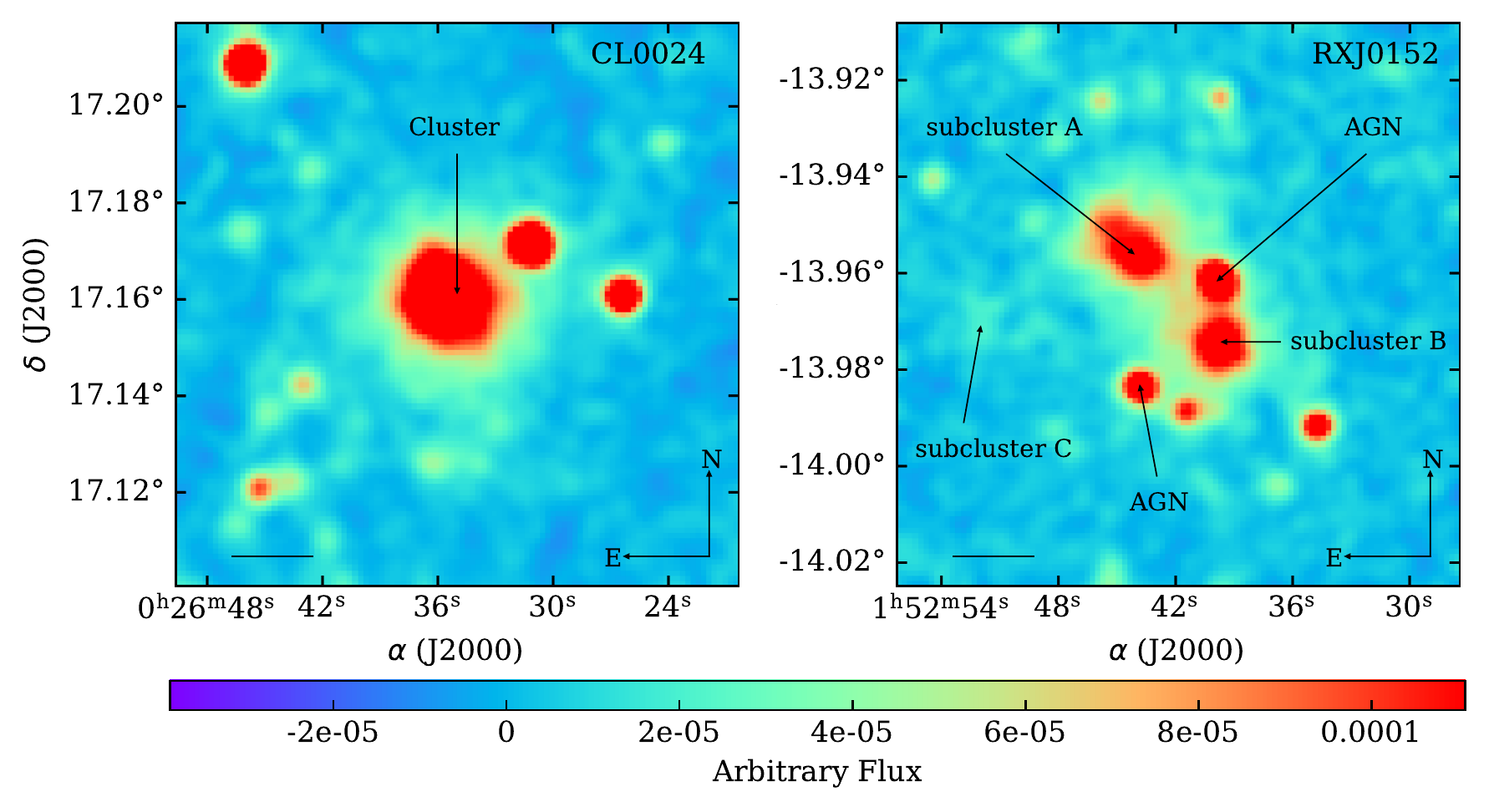}
\caption{Smoothed images of the X-ray emission detected with the \textit{XMM-Newton}. North is up and East is left. Images are $\sim7$ arcmin on the side corresponding to a physical size of 2.16 Mpc and 3.02 Mpc, for CL0024 at $z = 0.395$ (\textit{left panel}) and RXJ0152 at $z = 0.830$ (\textit{right panel}), respectively. The black solid bars at the bottom left of the panels show scales of 1 arcmin.}\label{fig:fig1}
\end{figure}

For data reduction, we used standard procedure using the XMM-Science Analysis Software (SAS) version 9.0.0 and the calibration database. Here we explain the main steps of our data reduction processes. The X-ray data consist of two types: observational data files (ODF) and current calibration files (CCF). The procedure to prepare the ODF is as follows. First, we produce a light curve from data in the high energy range to study the temporal variability due to non-target sources (i.e., instruments, Sun, and Milky Way). The X-ray emission from the galaxy cluster itself is not expected to be variable within the observation period. We used an energy range of 10 - 12 keV for the EMOS camera and 12 - 14 keV for EPIC pn camera. We applied 3$\sigma$ clipping to reject time intervals which could be affected by flares in the range of 0.3 - 10 keV. Second, we filtered event files using pattern and flag criteria to get good quality event files. In particular, to study the extended emission from the galaxy cluster we subtracted point source contamination using Chandra reported in \cite{Jee+07}. The final event files were then used for imaging and spectral analyses. The X-ray images of the two clusters shown in Figure \ref{fig:fig1} are Gaussian-smoothed with ${\sigma =7"}$, roughly $1\%$ of the clusters' width.

\begin{figure}[!htbp]
\centering
\includegraphics[width=\columnwidth]{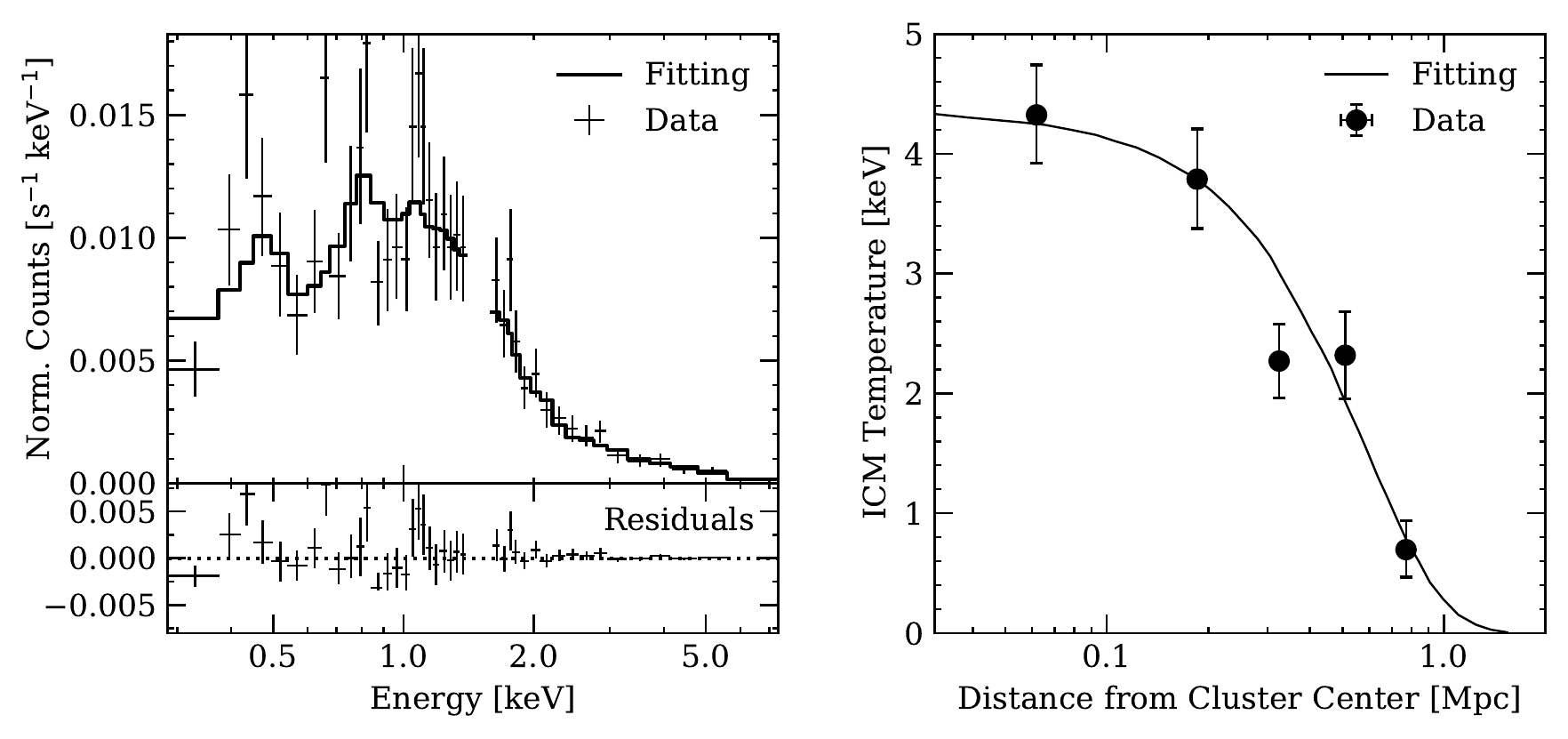}
\caption{Results from \textit{XMM-Newton} of CL0024: \textit{Left panel}: central part spectral energy distribution. \textit{Right panel}: Projected ICM temperature profile.}\label{fig:fig2}
\end{figure}

The ICM temperature is estimated by analyzing the spectra using \texttt{XSPEC}, an X-Ray Spectral Fitting Package. We fitted the filtered X-ray data with thermal plasma models using absorbed MEKAL model in energy range from 0.3 to 10 keV excluding 1.4 - 1.6 keV region due to Al profile and 7.45 - 9 keV region due to Cu line. The MEKAL model is used to describe emission spectrum of hot diffuse gas based on the physical model described in Mewe \textit{et al.} in \cite{Mewe+85,Mewe+86}, Kaastra in \cite{Kaastra+92}, and Liedahl \textit{et al.} in \cite{Liedahl+95}. There are five parameters in MEKAL model namely, plasma temperature (in units of keV), hydrogen density, metal abundance, redshift, and a normalization factor that one can choose to calculate or interpolate from a spectrum model. We combined this model with an absorption model following a local H{\sc i} observation which proposed by Dickey \& Lockman in \cite{Dickey+90}. The redshift data of each cluster are used to fix the values of physical parameters that depend on cosmological distance. The chemical abundances are in units of the solar values. The output of the spectral fitting is the temperature of the ICM. We show the fitting results for the center part of CL0024 in the left panel of Figure \ref{fig:fig2}. The vertical spread is large for the low energy region. The explanation for this is that at this energy the contrast against the background is low, resulting in low S/N. Points with high counts are also identified by Zhang \textit{et al.} in \cite{Zhang+05} who attributed them to the complex structure at the central region of the cluster.

Whenever possible we divided the gas distribution into a number of concentric regions and determined the temperature in each of those concentric rings to construct a radial profile of the temperature. The right panel of Figure \ref{fig:fig2} shows the resulting temperature radial profile of ICM for cluster CL0024. Unfortunately, we could not use this approach for RXJ0152 due to its irregular distribution, nor could we produce radial profile for each sub cluster as the low signal to noise spectra prevented us from dividing the regions further. Instead, we decided to use the temperature of each sub cluster to characterize the ICM temperature distribution. The sub-regions are clearly seen in Figure \ref{fig:fig1} which is in good agreement with Jee \textit{et al.} in \cite{Jee+07}.

\subsection{Photometry and Spectroscopy}
For the optical band, we benefited from the following data resources and previous works: optical photometric data \cite{Blakeslee+06,Czoske+01,Kodama+04}, spectroscopic redshift \cite{Demarco+05,Homeier+05,Jorgensen+05}, and the calculation of mass distribution data (from gravitational lensing analysis) \cite{Jee+05,Ota+04,Umetsu+10,Wagner+18}. We use  $B$ and $V$ bands for CL0024 and $r, i, z$ bands for RXJ0152 which is at a higher redshift. The photometric data are used for the color-magnitude diagrams we need to compensate the inadequate information on the morphology of the galaxies, whereas the spectroscopic data are used to calculate the SFR.

\section{Analysis}\label{sec:analysis}
\subsection{General features of the clusters}
Analysis of the redshift distribution of the member galaxies of CL0024 reveals sub clustering in velocity space, with populations of 200 and 303 galaxies in the redshift range of $0.374 \leq z < 0.387$ and $0.387 \leq z < 0.402$, respectively. Similar results are reported in previous works \cite{Czoske+01,Treu+03}. Gravitational lens analysis done by Umetsu \textit{et al.} in \cite{Umetsu+10} reveals a mass profile of two merging sub-clusters suggesting a total mass of $\sim 10^{15} h^{-1}M_{\odot}$ with $h=H_0/100$. X-ray emission of this cluster shows extended features with regular shape if we exclude the outer point sources (see the left panel of Figure \ref{fig:fig1}).

Cluster RXJ0152 is very rich in ICM which emits high luminosity X-ray in a non-axially-symmetric distribution, even though the galaxies are distributed mainly along the northeast–southwest direction. This cluster was examined by the Reionization Lensing Clusters Survey (RELICS). A thorough report by Acebron \textit{et al.} in \cite{Acebron+19} indicates strong lensing and suggesting an ongoing massive cluster strengthening with current estimation of the total mass is $10^{14} h^{-1}M_{\odot}$.

\subsection{Properties of the galaxies}
We make use of the information on the morphology of each galaxy given by Treu \textit{et al.} in \cite{Treu+03} for CL0024, and by Postman \textit{et al.} in \cite{Postman+05} for RXJ0152. The morphology of each cluster galaxy is determined by applying two classification schemes, i.e., the Medium Deep Survey scheme for CL0024 (see Table \ref{tab:tab2}) and the Hubble $T$-type for RXJ0152 (see Table \ref{tab:tab3}). Only about a half of the galaxies have assigned morphologies using these schemes which hinders our attempt to examine the original Dressler's morphology-relation. For all galaxies we use galaxy color-magnitude diagnostic to classify morphology of galaxies, albeit only broadly.

\begin{table}[!htbp]
\centering
\caption{Medium Deep Survey morphology classification scheme.}
\fontsize{9pt}{9pt}\selectfont
\begin{tabular}{ll}
\hline\T
\textbf{Sample}&\textbf{Morphology}\B\\
\hline\T
0 &	Elliptical (E) \\
1 &	Elliptical/Lenticular (E/S0) \\
2 &	Lenticular (S0) \\
3 &	Spiral Sa and Sb \\
4 &	Spiral \\
5 &	Spiral Sc and Sd \\
6 &	Irregular \\
7 &	Unclassified \\
8 &	Merger \\
\hline
\end{tabular}
\label{tab:tab2}
\end{table}

\begin{table}[!htbp]
\centering
\caption{Hubble $T$-type morphological classification scheme.}
\fontsize{9pt}{9pt}\selectfont
\begin{tabular}{ll}
\hline\T
\textbf{Class}&\textbf{Morphology}\B\\
\hline\T
$-5 \leq T \leq -3$ & Elliptical (E) \\
$-2 \leq T \leq 0$ & Lenticular (S0) \\
$1 \leq T \leq 8$ & Spiral (S) \\
10 & Irregular \\
 \hline
\end{tabular}
\label{tab:tab3}
\end{table}

Galaxy’s color represents the stellar constituent of the galaxy which could offer robust probe on cluster history. It is important to note that galaxy may be formed, deformed, and reformed, whereas a star generally goes along a one-way evolutionary track.

We use optical color for both clusters to measure the color of their galaxies. We use $B-V$ for CL0024 based on observation by the CFHT \cite{Czoske+01} and $r-z$ for RXJ0152 using the Advanced Camera for Survey (ACS) aboard the \textit{Hubble Space Telescope} \cite{Blakeslee+06}.

\begin{figure}[!htbp]
\centering
\includegraphics[width=\columnwidth]{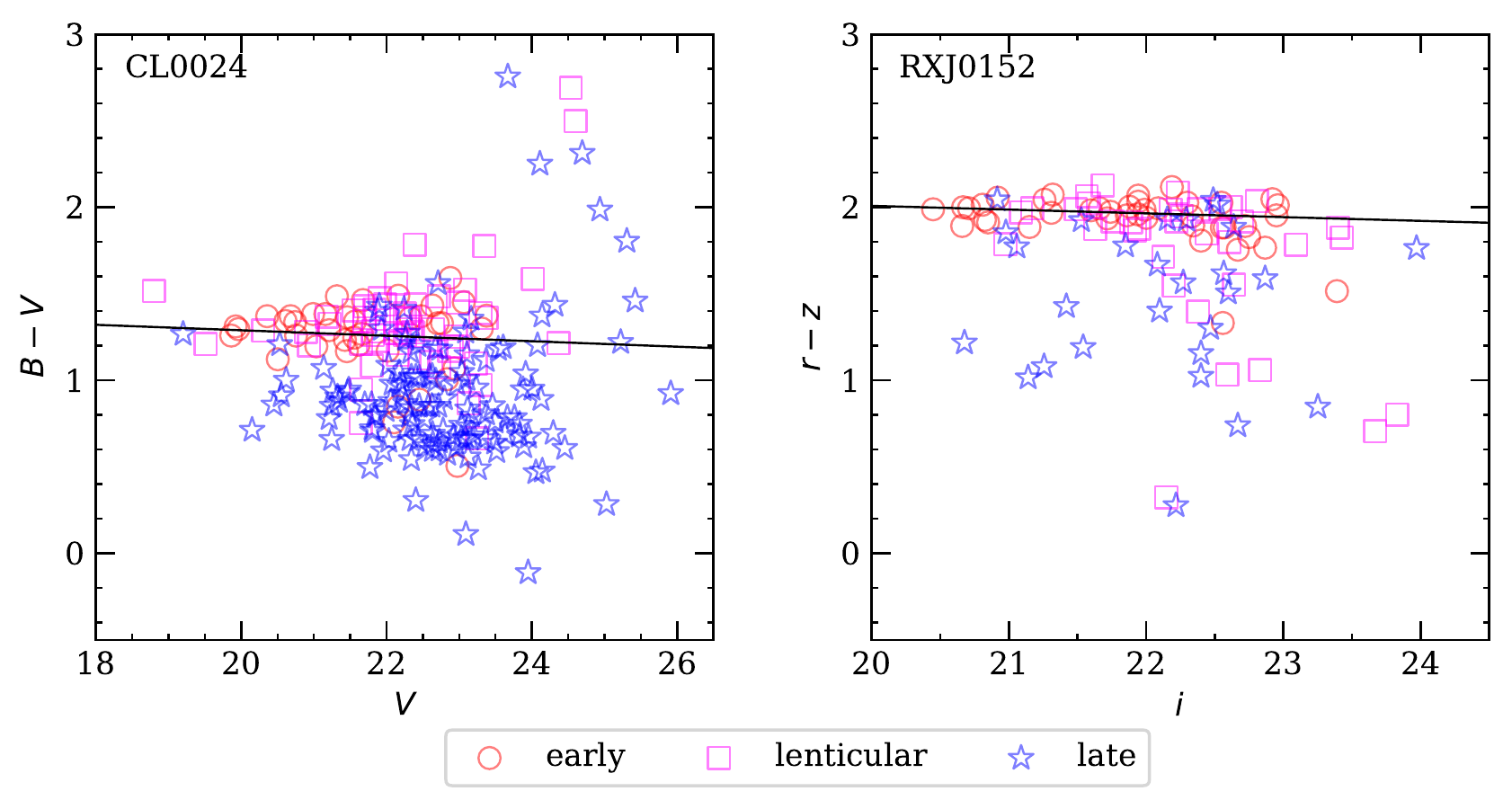}
\caption{CMD for CL0024 (\textit{left panel}) and RXJ0152 (\textit{right panel}). The symbols represent three types of morphology: circles, squares, and stars are for early-type, lenticular, and late-type galaxies, respectively. The solid black lines trace the red sequence.}
\label{fig:fig3}
\end{figure}

The color-magnitude diagrams (CMD) for each cluster are shown in Figure \ref{fig:fig3}. For both CMDs, early-type galaxies are mainly found along the so called narrow “red sequence” distribution, whereas lenticular and late-type galaxies are more spread-out.

For a better probe of physical condition of cluster members, we study the galaxies’ SFR using a chosen set of emission lines of the galaxy. SFR may be used as a proxy indicating the rate of cool gas consumption to form stars in a galaxy. Significant deviation from certain typical value of SFR of isolated galaxies might indicate non-secular galaxy evolution involving interaction between the galaxy and its environment.

Various star formation tracers exist in the literature, each with its advantages and disadvantages. For this study [O{\sc ii}] $\lambda$3727 \AA~ line is used in favor of more established tracers such as H$\alpha$. The reason is that for redshift higher than 0.5 H$\alpha$ emission appears in infrared domain where observation is more challenging than in the optical. For RXJ0152 we used SFR value reported by Homeier \textit{et al.} in \cite{Homeier+05}, whereas for CL0024 we computed SFR using the equivalent width of [O{\sc ii}] line reported by Czoske \textit{et al.} in \cite{Czoske+01} and Moran \textit{et al.} in \cite{Moran+05}. At the cluster redshift, [O{\sc ii}] line is shifted into $V$-band hence we could calculate [O{\sc ii}] line luminosity knowing the observed magnitude and the luminosity distance of each galaxy, using the following formula \cite{Hopkins+03,Jorgensen+05}:
\noindent 
\begin{equation}
    \mathrm{SFR_{[O \textsc{ii}]}} (M_{\odot}\mathrm{yr}^{-1})=\frac{L_{[\mathrm{O \textsc{ii}}]}}{2.97\times 10^{33}\mathrm{W}} \qquad {\rm for\ CL0024},
\end{equation}
\begin{equation}
    \mathrm{SFR_{[O \textsc{ii}]}} (M_{\odot}\mathrm{yr}^{-1})=6.58 \times 10^{-42}L_{[\mathrm{O \textsc{ii}}]} \qquad {\rm for\ RXJ1152}
\end{equation}
with
\begin{equation}
L_{[\mathrm{O \textsc{ii}}]}=3.11 \times 10^{-20}[L_{[\mathrm{O \textsc{ii}}]}]_{\rm obs}^{1.495},
\end{equation}
and  
\begin{equation}
[L_{[\mathrm{O \textsc{ii}}]}]_{\rm obs}=1.4 \times 10^{29}\frac{L_{\mathrm{B}}}{L_{\mathrm{B\odot}}} EW_{[\mathrm{O \textsc{ii}}]},
\end{equation}
where $L_{\mathrm{B}}/L_{\mathrm{B\odot}}=10^{0.4(5.48-M_B)}$ (see details in \cite{Balogh+97,Kennicutt+92}) and $M_B$ is absolute $B$ magnitude.

As our computation only took [O{\sc ii}] emission line in $V$-band, our resulting SFR might suffer methodological incompleteness, which will be improved by adding other lines in infrared and ultraviolet bands for the SFR computation input.

\subsection{Properties of the environment}
For both clusters, the immediate environment of each galaxy is characterized by the surface number density around it, and is studied using the galaxy position in the optical data. A galaxy is termed “the target galaxy” when its environment is being considered. All galaxies take turn to be considered as the target galaxy. The number density, $\Sigma_{10}$, of each environment is calculated within a circular area of radius $R_{10}$, which is the projected distance to the farthest galaxy among the nearest 10 neighbors of the target galaxy. We also calculate the physical distance of each target galaxy from the center of the cluster. We also study the gaseous environment of each galaxy by checking the temperature of the gas located at the same radial distance from the cluster center. This is done because the spatial resolution of the X-ray data has not allowed us to calculate the temperature of the gas immediately surrounding each galaxy. While locally inaccurate, the smooth temperature profile of the gas (Figure \ref{fig:fig2}) assures this approach gives representative cluster radial profile.
\begin{figure}[!htbp]
\centering
\includegraphics[width=\columnwidth]{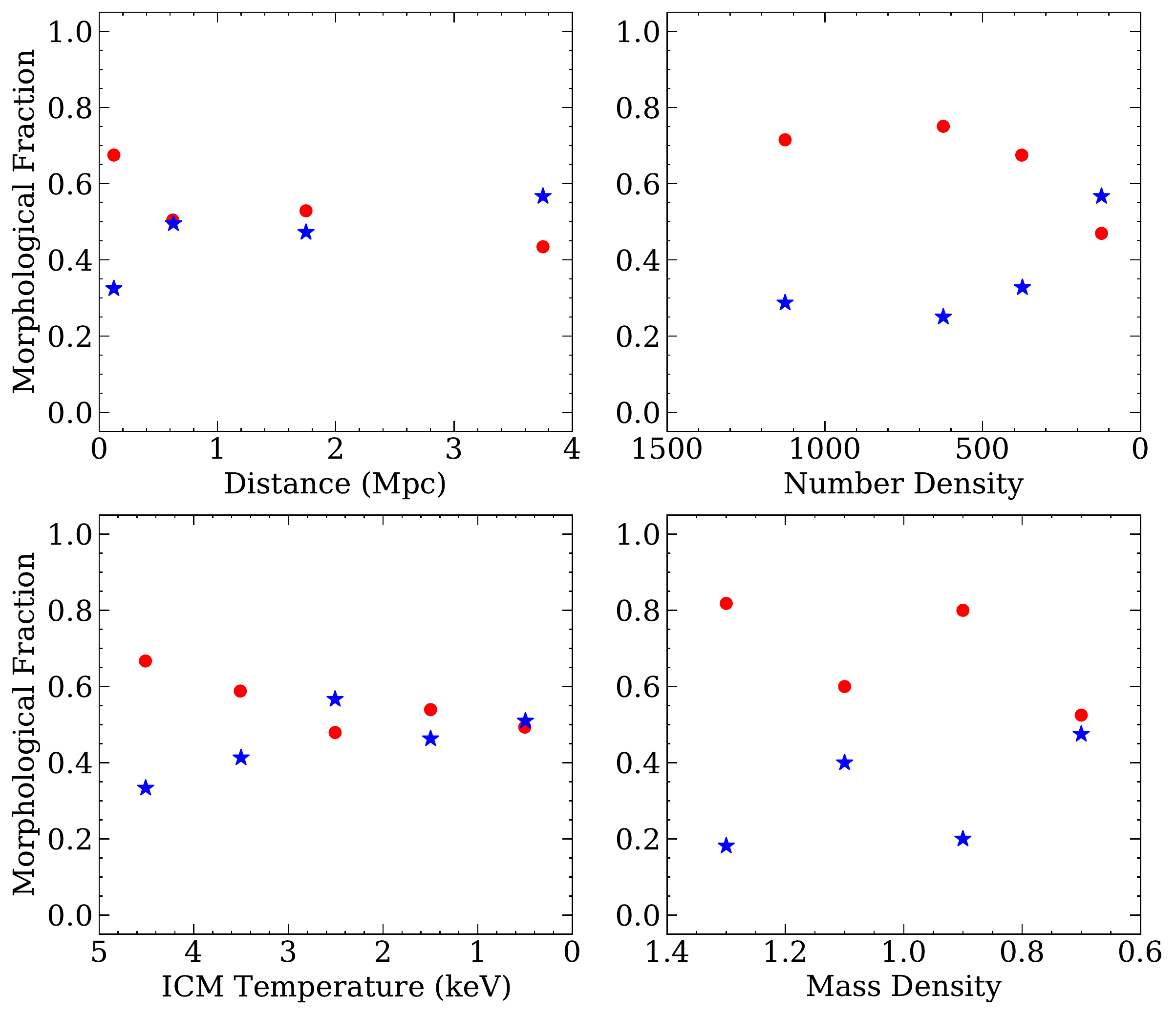}
\caption{Relation between the morphological fraction with environment properties (i.e., distance from the center of cluster, number density, ICM temperature, and  mass density) for CL0024. The symbols are the fraction value in each binning in the $x$-axis values which} denote early-type (circle) and late-type (star) galaxies. Morphological type refers to Table \ref{tab:tab2}.
\label{fig:fig4}
\end{figure}

\begin{figure}[!htbp]
\centering
\includegraphics[width=\columnwidth]{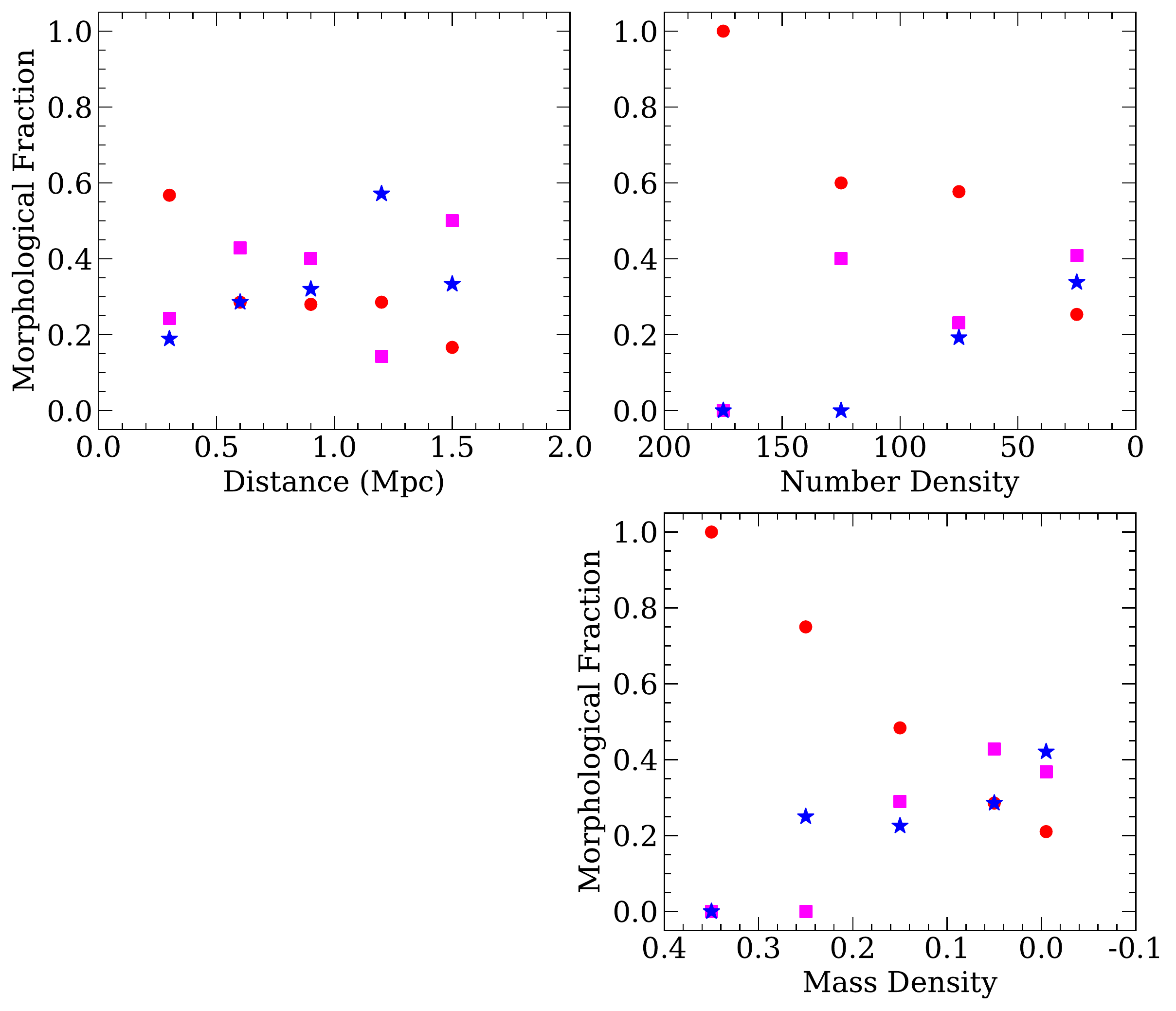}
\caption{Relation between the morphological fraction and environment properties for RXJ0152. The symbols are the fraction value in each binning in the $x$-axis values which} denote three morphological type, early-type (circle), lenticular (square), and late-type (star) galaxies. Morphological type refers to Table \ref{tab:tab3}.
\label{fig:fig5}
\end{figure}

We then evaluated the relationship between color, morphology, and SFR of each galaxy with its environment using Spearman’s rank correlation analysis, which although limited to linear correlation, is adequate for the purpose of recognizing broad trends in the correlations.

\section{Results and Discussion}\label{sec:results}
Once all the desired information on galaxy properties and environment properties are compiled, we study the statistical correlation between galaxy properties and the environments.

\subsection{Morphology}
Morphology-density and morphology-distance relations are the most classic examination in investigating the relation between galaxy property and its environment, pioneered by Dressler in \cite{Dressler+80}.

In Figure \ref{fig:fig4} we see that toward the clusters center of CL0024, the early-type galaxies tend to dominate the morphological distribution. Our results shows the same general trend found earlier by Dressler in \cite{Dressler+80} and the computational galaxy evolution by Martel \textit{et al.} in \cite{Martel+98}. However, our results at the outskirt of the clusters slightly differ from the similar studies of more relaxed clusters.

Similar trends are found in cluster RXJ0152 (Figure \ref{fig:fig5}), although we are not yet able to find relation between galaxy morphology  and the temperature of its surrounding ICM. Both clusters show good smooth correlation between morphology with number density. It could be due to the fact that it is directly related to the galaxy as number density is calculated around each galaxy. On the other hand, morphology correlation with galaxy location from the cluster center is weaker and not smooth. This may reflect the cluster's internal structure. The correlations between morphology and mass density and ICM temperature show good trends but less smooth, suggesting the role of position which implicitly used in mass and temperature calculations.

\begin{figure}[!hp]
\centering
\includegraphics[width=\columnwidth]{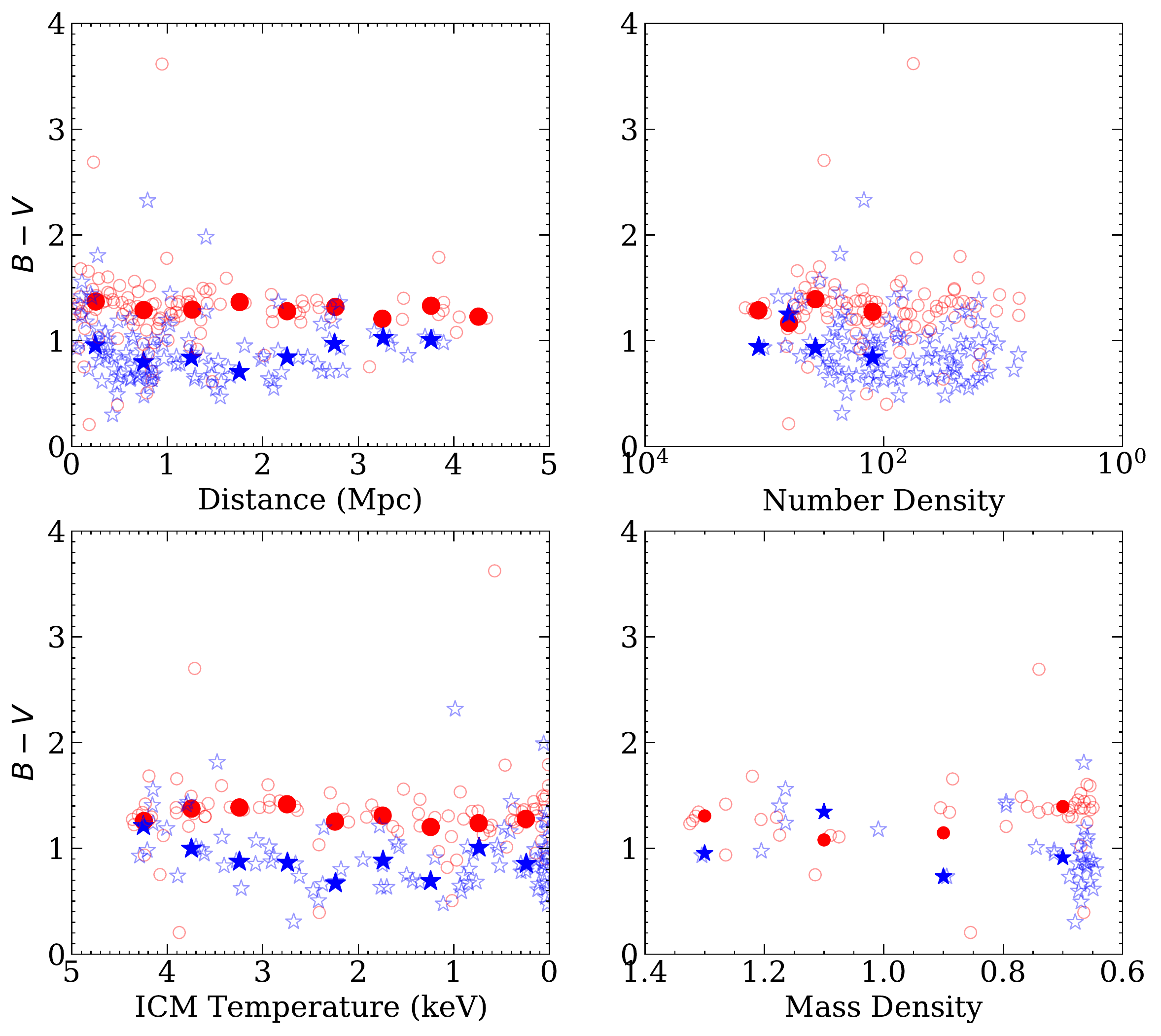}
\caption{Relation between $B-V$ color with environment properties for CL0024. The symbols denote two broad morphological type, early-type (circle) and late-type (star) galaxies. Filled symbols are the mean value in each binning in the $x$-axis values.}
\label{fig:fig6}
\end{figure}

\subsection{Color}

The color bimodality commonly exhibited by field galaxies \cite{Baldry+06} is also observed in our sample, especially in CL0024. It is shown in four panels of Figure \ref{fig:fig6}. The member galaxies categorized into two distinct types: red galaxies and blue galaxies. We find that the distribution of galaxy color in CL0024 does not vary significantly from the center to the outer region.

\begin{figure}[!hp]
\centering
\includegraphics[width=\columnwidth]{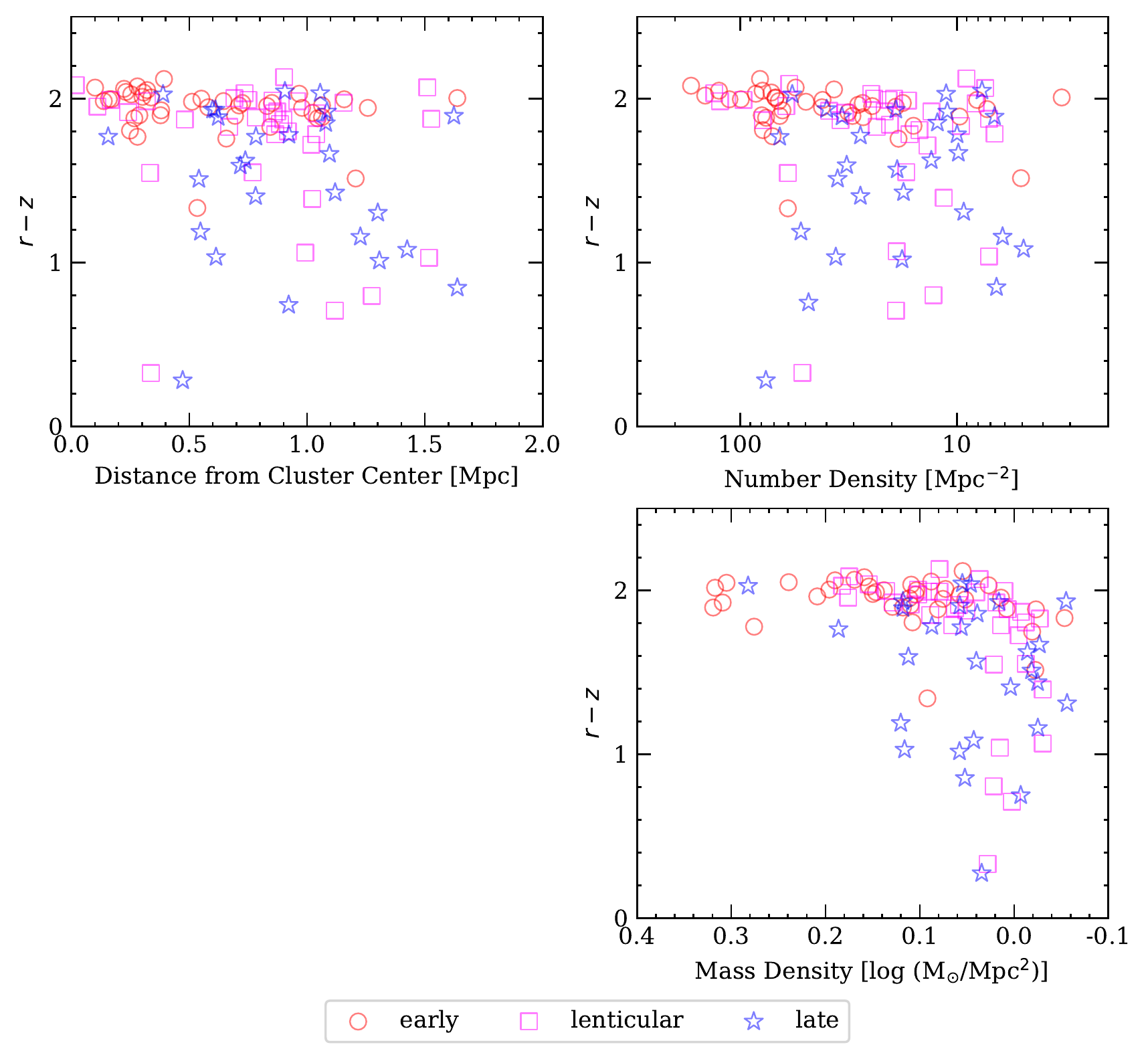}
\caption{Relation between $r-z$ color with environment properties for RXJ0152. The symbols denote three morphological types, early-type (circle), lenticular (square), and late-type (star) galaxies.}
\label{fig:fig7}
\end{figure}

The color distribution in RXJ0152 (see Figure \ref{fig:fig7}) is different from CL0024. In the central part, color of galaxies of RXJ0152 are distributed within narrow region, whereas it is more scattered in  the outer region.

\subsection{Star Formation Rate}
\begin{figure}[!htbp]
\centering
\includegraphics[width=\columnwidth]{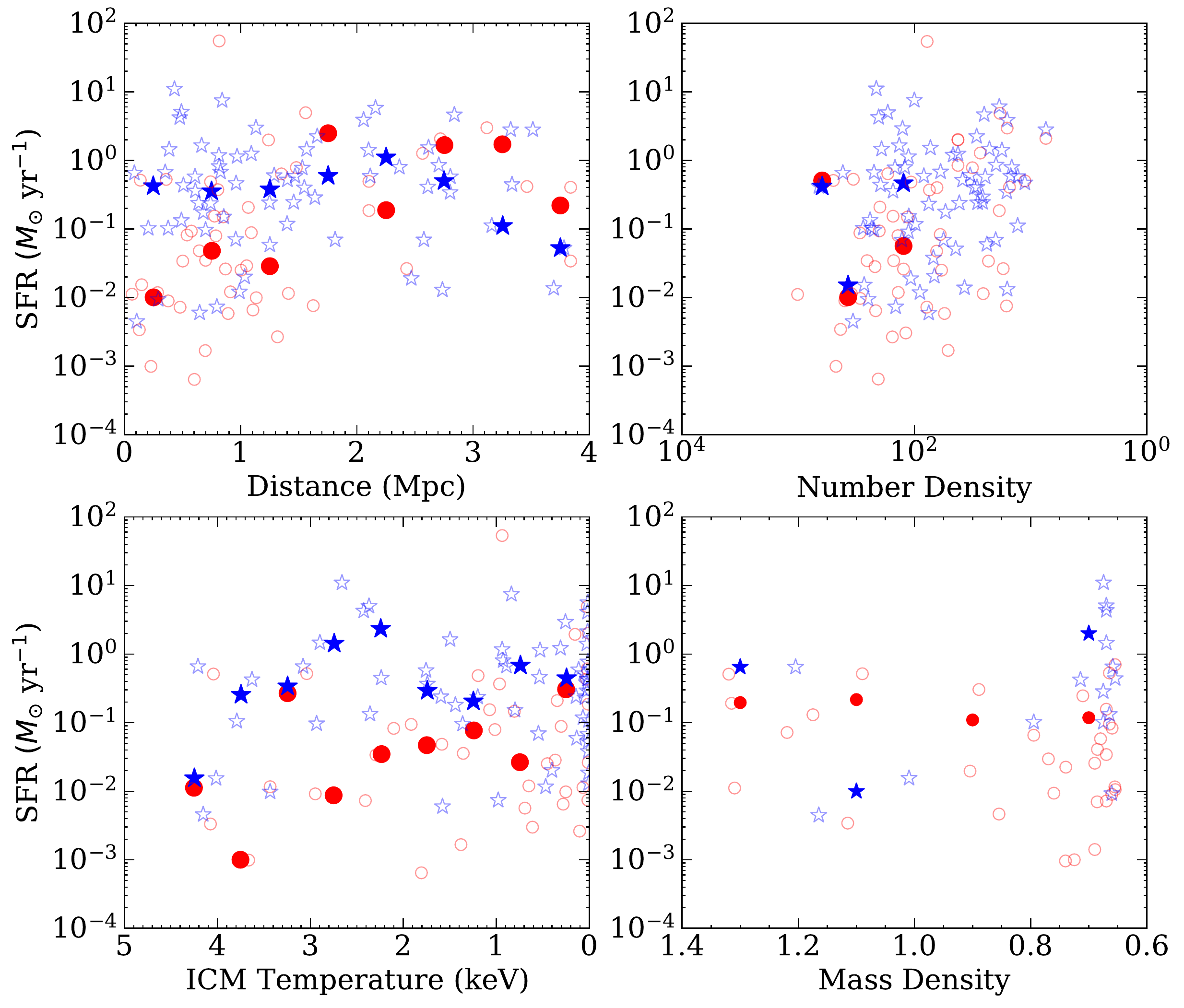}
\caption{Relation between the SFR with environment properties for CL0024. The symbols denote two broad morphological type, early-type (circle) and late-type (star) galaxies. Filled symbols are the mean value in each binning in the x-axis values.}
\label{fig:fig8}
\end{figure}

SFR can be used as a measure in classifying galaxies into two classes: red galaxies with no or little star formation and blue active star forming galaxies. In CL0024 we see SFR increases from outer region inward to radius 2.5 Mpc from the center and then decreases toward the center (see upper left panel of Figure \ref{fig:fig8}). If we use only two bins from the plot of SFR and number density, we see clear anti correlation of SFR and number density. SFR is increasing for galaxies reside on low density environments. ICM density can also affect cold gas supply inside of the galaxies that fuels star formation. The lower left panel of Figure \ref{fig:fig8} shows SFR decreases with increasing ICM temperature. Combining this trend with the X-ray emission distribution shown in Figure \ref{fig:fig1} and the temperature profile shown in Figure \ref{fig:fig2} suggests SFR decreases toward the center where the peak of the hot ICM distribution located.

\begin{figure}[!htbp]
\centering
\includegraphics[width=\columnwidth]{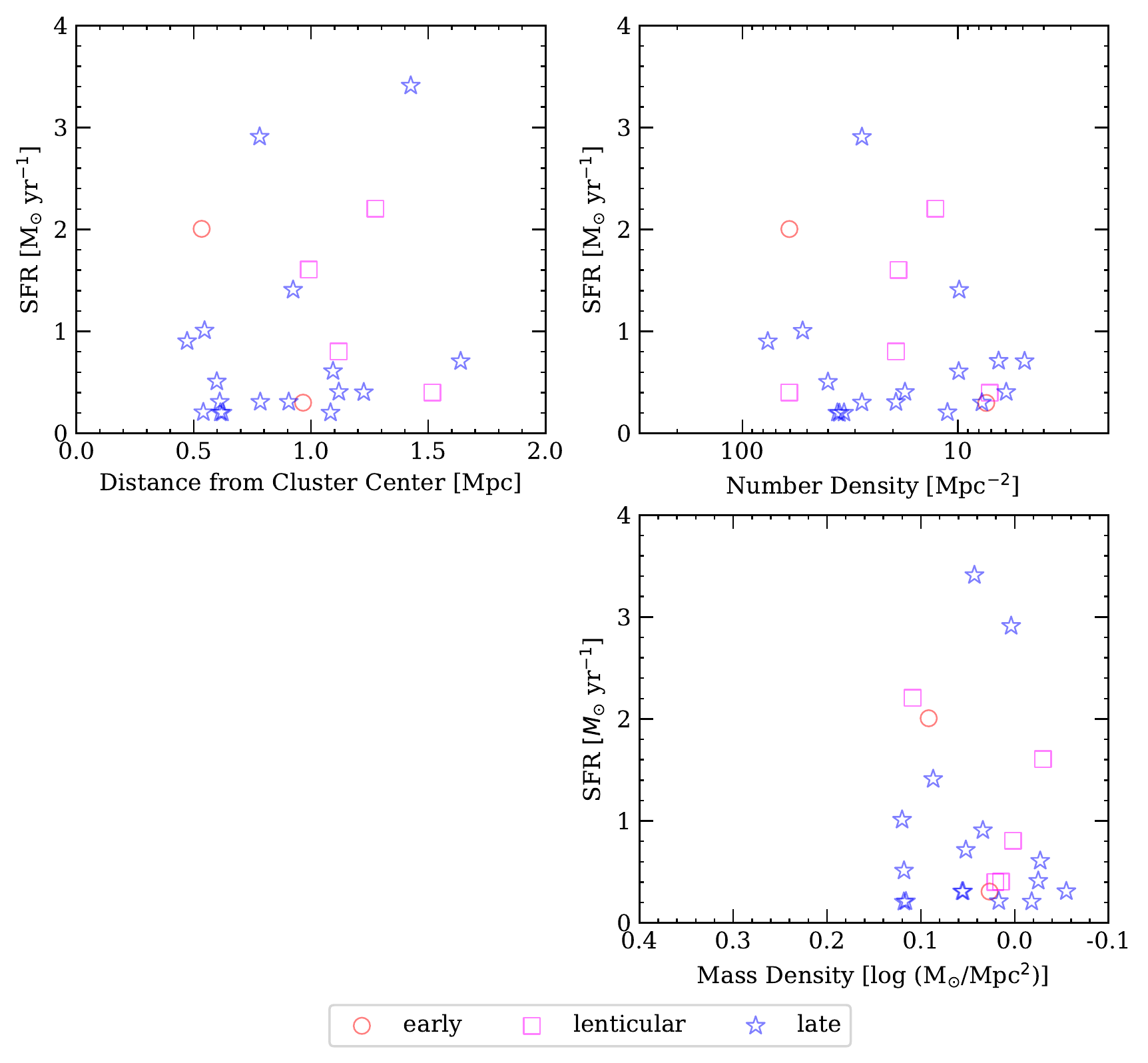}
\caption{Relation between the SFR with environment properties for RXJ0152. The symbols denote three morphological types, early-type (circle), lenticular (square), and late-type (star) galaxies.}
\label{fig:fig9}
\end{figure}

\begin{figure}[!htbp]
\centering
\includegraphics[width=\columnwidth]{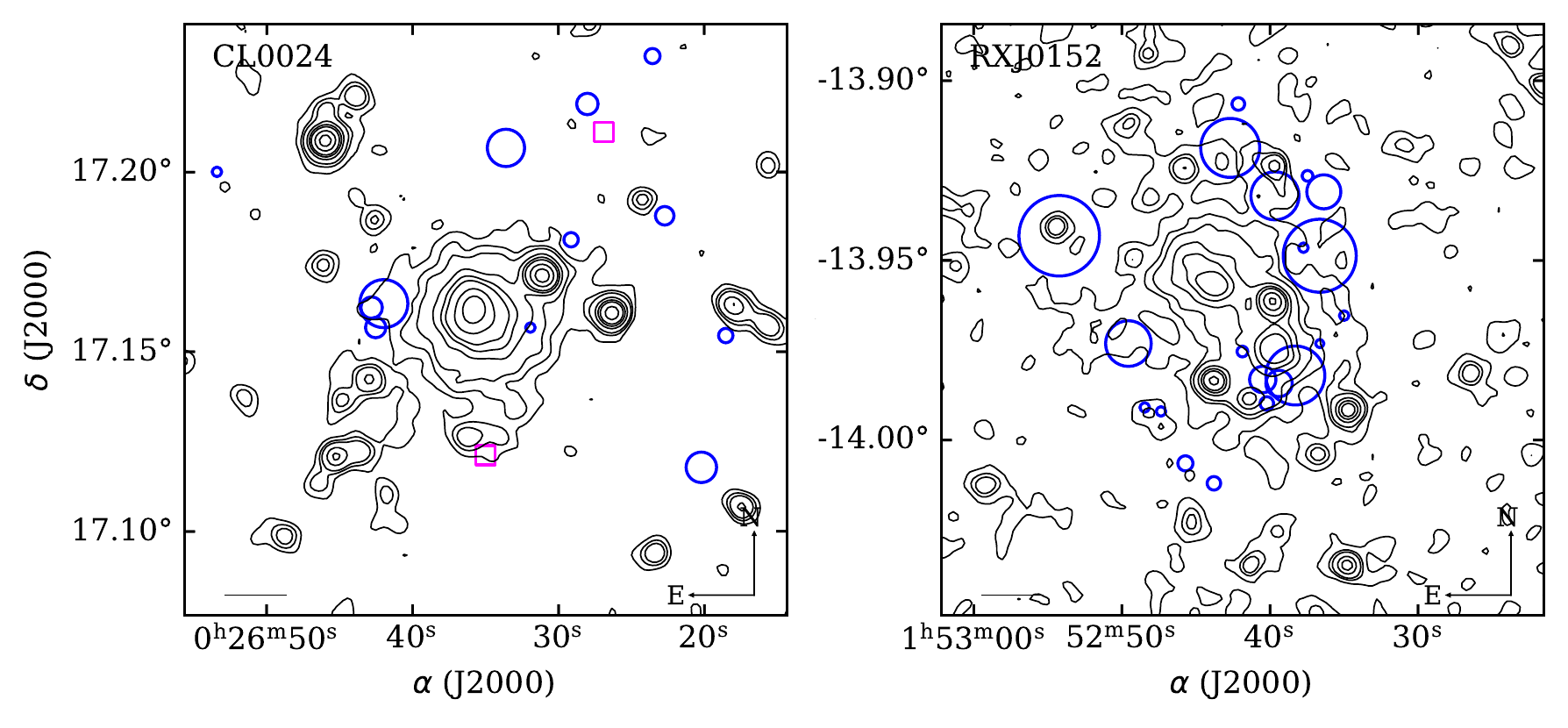}
\caption{Spatial distribution of star forming galaxies (indicated by circles) in reference to the ICM distribution in CL0024. The size of the circle is proportional to the SFR. The two galaxies with high SFR ($> 50 M_{\odot}$ yr$^{-1}$) are indicated by box symbol. The 1 arcmin bar corresponds with 0.31 Mpc at the cluster’s redshift. Spatial distribution of star forming galaxies (indicated by circles) with respect to ICM distribution in RXJ0152. The size of the circle is proportional to the magnitude of the SFR. North is up and East is left. The 1 arcmin bar corresponds with 0.43 Mpc at the cluster’s redshift.}
\label{fig:fig10}
\end{figure}

SFRs of cluster members for RXJ0152 do not build a clear trend with the distance to the center, number density, nor mass density. Our analysis suggests the role of the ICM in the process that would determine the properties of the cluster members. We study this by examining any possible correlation between the two. We show cluster members distribution with respect to ICM distribution of each cluster in Figure \ref{fig:fig9}. Upon close examination of Figure \ref{fig:fig10} one finds higher occurrence of star formation at less dense area surrounded by cooler ICM, with galaxies showing the strongest SFR residing at the periphery of the cluster.

\begin{table}[!htbp]
\centering
\caption{Spearman’s rank correlation coefficient.}
\fontsize{9pt}{9pt}\selectfont
\begin{tabular}{lrrr}
\hline\T
\textbf{Environment Properties} & \multicolumn{1}{c}{\textbf{$B-V$}} & \multicolumn{1}{c}{\textbf{Morphology}} & \multicolumn{1}{c}{\textbf{SFR}}\B\\
\hline\T
\textbf{CL0024} & & &\\
\quad Distances &	$-0.188\ (> 99\%)$ & $0.129\ (> 99\%)$ & $0.292\ (> 99\%)$ \\
\quad Number Density	& $0.263\ (> 99\%)$ &	$-0.209\ (> 99\%)$ &	$-0.460\ (> 99\%)$ \\
\quad ICM Temperature	& $0.182\ (95 - 98\%)$	& $-0.077\ (< 90\%)$ & $-0.164\ (90 - 95\%)$ \\
\quad Mass Density	& $0.135\ (< 90\%)$ & $-0.219\ (95 - 98\%)$	& $-0.159\ (< 90\%)$ \\
\T \textbf{RXJ0152} & & &\B\\
\quad Distances & $-0.290\ (> 99\%)$ & $0.267\ (> 99\%)$ & $0.137\ (< 90\%)$ \\
\quad Number Density	& $0.266\ (> 99\%)$ & $-0.346\ (> 99\%)$ & $-0.115\ (< 90\%)$ \\
\quad Mass Density	& $0.502\ (> 99\%)$ & $0.341\ (> 99\%)$ &	$0.126\ (< 90\%)$ \\
 \hline
\end{tabular}
\label{tab:tab4}
\end{table}

All relations are then quantified using Spearman's ranking scheme, which is adequate for this work, with the resulting correlation coefficients given in Table \ref{tab:tab4}. The distance, number density, ICM temperature, and mass density are listed from low to high; color \textit{B-V} from blue to red; morphology from early type to late type; and SFR from low to high. The numbers in parentheses are the levels of confidence. The Spearman's rank correlation coefficients show that the correlation between galaxies' properties and their environments are weaker compared to those of more relaxed clusters.  Nonetheless, they seem to follow the same direction as indicated by general evolution of bound large scale structure such as clusters (see, e.g., {\cite{Beyoro-Amado+21}}). The Dressler's morphology-density relation in our work is in the range of 0.209 - 0.346, weaker than, yet still overlaps at the lower end with the range of 0.271 - 0.96 found by Houghton in \cite{Houghton+15}. We redid the correlation analysis separately for each type of galaxy and found that the strength of the correlations differs for the two types (shown in Figure \ref{fig:fig3}). It indicates that the environmental mechanisms may operate differently for early and late-type galaxies or operate at different stages of their evolutions.

\subsection{Galaxy interaction with diffuse environment}

If we take a simple blackbody description of radiation, then the radiation pressure is proportional to the fourth power of temperature. The most interesting result found in our analysis on the correlation between SFR and ICM temperature (see Figure \ref{fig:fig8}) is that ICM pressure seems able to enhance the SFR of the galaxies near the virial radius, but to decrease the SFR for galaxies towards the cluster core. Environmental mechanisms that could give rise to such effects are ram pressure stripping and starvation by the ICM, and/or tidal compression by the cluster potential. The information is entangled that it is nearly impossible to single out any particular mechanism. However, in this work our attempt is restricted to evaluate the impact of only ICM's property on SFR. This preliminary directs us to possible parameters closely describe the ICM-galaxy interaction.

In the process of virialization, galaxies move about and inward in the cluster, and experience various physical processes along their courses. A galaxy moving towards and through a bulk of ICM would experience cluster potential gradient and also pressure exerted by the ICM. Depending on how the pressure is balanced by the galaxy potential, a portion of cool gas might be stripped of the galaxy. The incoming pressure depends on the velocity with which the galaxy encounters the ICM, the density of the ICM, and the cross section of the encounter. The correlation between ICM and galaxy property shown in Table \ref{tab:tab4} suggests a positive interaction in CL0024. However, at this stage we could not yet determine the relative contribution from each detailed mechanism. It is also imperative to recalculate the SFR using complete indicators to establish a finer correlation between the physical properties of galaxies and their immediate environment which represents the cluster stage of evolution.

The weak correlation between SFR and cluster property shown in Figures \ref{fig:fig8} and \ref{fig:fig9} are in agreement with previous studies on the relation between SFR and/or emission line strength have been studied by Balogh \textit{et al.} in \cite{Balogh+04} and Kodama \textit{et al.} in \cite{Kodama+04}. They interpret the weak correlation between those two parameters and local environment as weak dependence of galaxy on their current environment. The current (observed) condition of each galaxy is steered by its previous environment, thus unobserved. Instead we need to find correlation between the physical and environment parameters interpreted from Star Formation History.

\section{Summary and Conclusions}\label{sec:summary}
We have performed combined analyses of X-ray and optical data of two well-known galaxy clusters to examine possible correlation between the physical properties of galaxy members (i.e., morphological type, color, and SFR) with certain measures of the clusters' overall profiles: galaxy distance to the cluster center, and mass density, and also with its immediate environment: number density, ICM temperature. For both clusters good confidence level is seen on the correlation between galaxy's color and morphology with distance to the center, number density, and mass density. Smooth X-ray emission profile provides guidance to distribution of ICM, the more diffuse environment of each galaxy. It is characterized by its temperature and the flux of its X-ray emission.  In CL0024 SFR is well correlated with distance to the cluster center and number density, but less so with ICM temperature and mass density, which is quite typical among massive clusters with central mass dominance. In the younger cluster (RXJ0152), with mass  of one order of magnitude lower than CL0024, we see less confidence on the correlation between galaxy properties with its environment. The correlations between number density and all galaxy parameters (color, morphology, and SFR) in both clusters are the strongest and most consistent with absolute correlation coefficient in the range 0.115-0.460, with error estimates better than $5\%$. Correlation between mass density with color is three times stronger in RXJ0152 than in CL0024. However correlation coefficients between mass density and morphology, and also SFR, of the two clusters are of opposite sign, with error estimate of less than $10\%$ The correlation between galaxy physical properties with the cluster overall feature is weaker than with the galaxy local environmental properties. This indicates that local physical processes play more important role in galaxy evolution, even in a self-contained system such as a cluster. Moreover, cluster are prone to further aggregation causing it to take a long time to virialize, with the consequence galaxy and its immediate environment subjected to the cluster's detail structure while following virialization process. It is worth mentioning that the correlation is examined between averaged measures, which reflects on the non-smooth correlation due to small number of galaxies included in the statistics. Albeit preliminary, this result opens way for our further attempt to better understand how galaxy's environment dictates its course of evolution benefiting the future galaxy surveys which will reveal more homogeneous data of details.

\section*{Acknowledgement}
This research has made use of the NASA/IPAC Extragalactic Database (NED) which is operated by the Jet Propulsion Laboratory, California Institute of Technology, under contract with the National Aeronautics and Space Administration; the VizieR catalogue access tool, CDS, Strasbourg, France; NASA’s Astrophysics Data System; and data from the High Energy Astrophysics Science Archive Research Center (HEASARC), provided by NASA’s Goddard Space Flight Center based on observations obtained with \textit{XMM-Newton}, an ESA science mission with instruments and contributions directly funded by ESA Member States and NASA. The authors thank Dr. Sean M. Moran and Dr. John Blakeslee for sharing the data. PWP gratefully acknowledges the WCU Research Grant 2019 from Institut Teknologi Bandung, the insightful discussion with Professor K. T. Inoue and his hospitality at Kindai University where this work is finalized. Financial support from ITB Voucher scholarship to DHN is gratefully acknowledged. ATJ is supported by Japan Society for the Promotion of Science (JSPS) KAKENHI Grant Number JP17H02868.  We thank the referees whose comments help us in improving this paper.


\end{document}